\documentclass[journal=ancac3,manuscript=article]{achemso}
\usepackage[T1]{fontenc}
\usepackage{lmodern}
\usepackage{graphicx}
\usepackage[english]{babel}
\usepackage{textcomp}
\usepackage{amsmath}

\author{Alice Cartoceti}
\affiliation{Department of Energy, Politecnico di Milano, via Lambruschini 6, 20156, Milano, Italy}
\email{alice.cartoceti@polimi.it}

\author{Paolo D'Agosta}
\affiliation{Department of Energy, Politecnico di Milano, via Lambruschini 6, 20156, Milano, Italy}

\author{Francesco Tumino}
\affiliation{Department of Energy, Politecnico di Milano, via Lambruschini 6, 20156, Milano, Italy}

\author{Valeria Russo}
\affiliation{Department of Energy, Politecnico di Milano, via Lambruschini 6, 20156, Milano, Italy}

\author{Carlo S. Casari}
\affiliation{Department of Energy, Politecnico di Milano, via Lambruschini 6, 20156, Milano, Italy}

\author{Andrea Li Bassi}
\affiliation{Department of Energy, Politecnico di Milano, via Lambruschini 6, 20156, Milano, Italy}
\email{andrea.libassi@polimi.it}

\title{Multiwavelength Raman investigation of\\mono- and few-layer MoS$_2$ grown by Pulsed Laser Deposition on SiO$_2$}

\begin{document}
\maketitle
\begin{abstract}
Molybdenum disulfide (MoS$_2$) is a semiconducting material whose vibrational and excitonic properties are highly sensitive to layer number and structural disorder. In this work, we demonstrate the possibility of producing MoS$_2$ monolayers on inert, electronics-compatible SiO$_2$ substrates via a room-temperature pulsed laser deposition (PLD) approach. The fine controllability of the process parameters allows tuning the number of deposited layers from mono- to multilayer, and these systems have been systematically investigated using multiwavelength Raman spectroscopy. The evolution of the Raman shift difference between the E\textsubscript{2g}\textsuperscript{1} and A\textsubscript{1g} modes, combined with the assessment of defect density, enables tracking of film growth as a function of the number of laser pulses during deposition. While excitonic effects are known to strongly influence the optical response of two-dimensional transition metal dichalcogenides, leading to resonance effects, experimental reports showing symmetry-selective exciton–phonon coupling have so far been limited. Here, we provide experimental evidence of symmetry-dependent exciton–phonon coupling in PLD-grown monolayer MoS$_2$. Specifically, we observe a modulation in the resonant behaviour of the out-of-plane (A\textsubscript{1g}) and in-plane (E\textsubscript{2g}\textsuperscript{1}) modes, which is related to their different coupling to the A excitons, predominantly derived from Mo\textsubscript{d$_{z^2}$} orbitals, and the C excitons, characterized by a mixed orbital character involving both Mo\textsubscript{d$_{z^2}$} orbitals and S p\textsubscript{x} and p\textsubscript{y} states. A direct comparison with mechanically exfoliated monolayers reveals the critical role of growth-induced defectivity in modulating these interactions. 
These findings establish a room-temperature PLD-based approach as a viable strategy for the growth of two-dimensional MoS$_2$ on inert, electronics-compatible substrates, and provide new insight into the interplay between excitonic resonances and growth-induced disorder in two-dimensional MoS$_2$. 
		
\end{abstract}
\vspace{5mm}

\section*{Introduction}
Molybdenum disulfide (MoS$_2$) has emerged as one of the most intensively investigated transition metal dichalcogenides (TMDs) due to its remarkable electronic, optical, and vibrational properties. In its most stable phase, 2H-MoS$_2$ consists of stacked S–Mo–S layers arranged in a trigonal prismatic coordination. While bulk MoS$_2$ is an indirect-gap semiconductor with a band gap of approximately 1.2 eV, reducing the thickness to a single MoS$_2$ layer induces a transition to a direct band gap in the visible range ($\sim$1.8–1.9~eV), resulting in a pronounced enhancement of photoluminescence (PL). Thanks to their carrier mobility, strong optical absorption, and high current on/off ratios in field-effect transistors, monolayer and few-layer MoS$_2$ are promising candidates for low-power electronics, photodetectors, light-emitting devices, and flexible optoelectronics \cite{eda2011photoluminescence,radisavljevic2011single,feng2012strain,he2013experimental,hui2013exceptional,mak2010atomically, li2012fabrication,wang2012electronics,ganatra2014few,cao2021roadmap,choi2017recent}.

Despite these appealing characteristics, the scalable fabrication of large-area, uniform, and high-crystalline-quality MoS$_2$ films remains challenging. Top-down approaches, such as mechanical or chemical exfoliation, provide high-quality flakes but suffer from poor scalability, random flake dimensions, and limited thickness control. Bottom-up techniques, such as chemical vapour deposition (CVD), enable the growth of large-area single- and few-layer MoS$_2$ as well as van der Waals heterostructures, but typically require high growth temperatures (1020 K up to 1270 K) and often suffer from limited reproducibility \cite{wang2012electronics,liu2012growth,xu2022growth,mandyam2020large}.

These limitations have motivated the exploration of alternative synthesis routes combining scalability, thickness control, moderate thermal budget and substrate versatility. Pulsed laser deposition (PLD) is particularly attractive in this respect, as it enables the stoichiometric transfer from a solid target, precise thickness control, and compatibility with a wide range of substrates \cite{late2014pulsed,serna2016large,serrao2015highly,seo2018growth,siegel2015growth,sinha2021study}. However, in conventional high-temperature PLD, MoS$_2$ is deposited while keeping the substrate at a high temperature ($\sim$970 K), determining chalcogen deficiency, vacancy formation, and oxidation of the deposited films, with potential degrading effects on their optical and electronic properties \cite{serna2016large,fominski2019formation,xie2019annealing,bertoldo2021intrinsic}. Previous reports from our group on the growth of MoS$_2$ on noble metals \cite{tumino2019pulsed,tumino2021hydrophilic} demonstrated that it is possible to exploit a PLD-based approach in which amorphous MoS$_2$ is deposited on the substrate kept at room temperature, and then it is crystallized by post-deposition annealing at about 700 K. More recently, another report exploited the same technique to produce 2-to-10-layer MoS$_2$ films on SiO$_2$ surfaces \cite{jaiswal2023wafer}. This strategy allows preservation of MoS$_2$ stoichiometry while enabling structural reconstruction into polycrystalline 2D layers, and it relies on a more accessible setup than the one required for high-temperature PLD. However, the production of MoS$_2$ monolayer on SiO$_2$ through room-temperature PLD is still unreported.

Herein, we prove the possibility to obtain a fine control over the deposited number of MoS$_2$ layers, down to the monolayer, on electronics-friendly SiO$_2$ substrates, through the abovementioned room temperature PLD-based approach.

In this respect, Raman spectroscopy represents a powerful, non-destructive tool for probing the structural and vibrational properties of bulk and low-dimensional MoS$_2$. In bulk 2H-MoS$_2$, two prominent first-order Raman-active modes are observed: the in-plane E\textsubscript{2g}\textsuperscript{1} mode ($\sim$383 $cm^{-1}$), involving opposite vibrations of sulfur atoms relative to the molybdenum atom, and the out-of-plane A\textsubscript{1g} mode ($\sim$408~$cm^{-1}$), associated with opposite vibrations of sulfur atoms perpendicular to the basal plane \cite{zhang2015phonon,zeng2012low}. On weakly-interacting surfaces, as the thickness decreases from multilayer to monolayer, the E\textsubscript{2g}\textsuperscript{1} mode upshifts, while the A\textsubscript{1g} mode downshifts. This behaviour originates from the competition between long-range Coulomb interactions, strongly affected by dielectric screening, and short-range interlayer forces \cite{chakraborty2013layer,molina2011phonons}. The resulting Raman shift difference $\Delta\omega$ between the two modes decreases monotonically with thickness, providing a reliable metric for layer-number determination when substrate
interactions are weak. For mechanically exfoliated MoS$_2$ on SiO$_2$, $\Delta\omega$ decreases from $\sim$26~$cm^{-1}$ in multilayer, bulk-like, samples to $\sim$19~$cm^{-1}$ in monolayers \cite{chakraborty2013layer,placidi2015multiwavelength,molina2011phonons,li2012bulk,lee2010anomalous,ottaviano2017mechanical}, while slightly different values are typically observed in the counterpart high-temperature PLD-grown films, i.e., $\sim$25~$cm^{-1}$ in the multilayer, bulk-like, configuration and $\sim$20.5~$cm^{-1}$ in monolayer MoS$_2$ \cite{serna2016large,serrao2015highly,sinha2021study}. In ideal, defect-free MoS$_2$, Raman scattering is restricted to zone-center phonons due to momentum conservation, i.e., E\textsubscript{2g}\textsuperscript{1} and  A\textsubscript{1g} Raman active modes. However, structural disorder breaks translational symmetry and relaxes Raman selection rules, enabling the activation of phonons with finite wavevector through defect-assisted scattering. A prominent disorder-activated feature is the longitudinal acoustic phonon at the M point, LA(M) ($\sim$227~$cm^{-1}$), which can be exploited to estimate the average interdefect distance in low-dimensional MoS$_2$ \cite{mignuzzi2015effect,stanford2019lithographically}.

Photoluminescence spectroscopy complements Raman measurements by providing direct insight into excitonic transitions, recombination dynamics, and electron–phonon interactions. In monolayer MoS$_2$, the conduction-band minimum and valence-band maximum are located at the inequivalent K points of the Brillouin zone, and optical excitation predominantly generates tightly bound excitons due to the strong Coulomb interaction arising from reduced dielectric screening in the two-dimensional limit. Two prominent excitonic features, labeled A and B, appear between $\sim$~1.8 and 2.1~eV and originate from direct optical transitions at the K point involving the lowest conduction band and the spin–orbit–split valence bands. The energy separation between the A and B excitons reflects the strong spin–orbit coupling of the valence band. In addition to these states, higher-energy excitonic resonances, commonly referred to as C excitons, have been identified, giving rise to a broad absorption band centred around 2.7~eV \cite{mak2010atomically,placidi2015multiwavelength,coehoorn1987electronic,livneh2010resonant,mccreary2018and,xiao2012coupled,burns2020controlling,carvalho2016erratum,scheuschner2012resonant,ellis2011indirect}. While A and B excitons predominantly derive from Mo\textsubscript{d$_{z^2}$} orbitals, exhibiting azimuthally symmetric wavefunctions, C excitons are instead characterized by a mixed orbital character involving both Mo\textsubscript{d$_{z^2}$} orbitals and S p\textsubscript{x} and p\textsubscript{y} states. When monolayer MoS$_2$ is excited at energies that are close, or directly match, the abovementioned absorption bands, resonant Raman effects can take place, allowing for the obtainment of further insights into the optoelectronic properties of the system.

Exploiting the complementarity of Raman spectroscopy and photoluminescence, we report on the multiwavelength Raman investigation of the mono- to multilayer MoS$_2$ samples obtained via room-temperature PLD on SiO$_2$. Through the combination of resonance Raman effects, photoluminescence spectroscopy and scanning electron microscopy, we unveil a symmetry-dependent exciton-phonon coupling in MoS$_2$ monolayer, which turns out to be affected by the system defectiveness.

\section*{Methods}
Oxidized p-doped Si substrates with 285~nm thick thermally grown oxide (SiO$_2$) were firstly subjected to a double-step cleaning cycle via sequential rinsing into acetone and methanol, each followed by rinsing in deionized water, then dried by fluxing nitrogen gas. Finally, they were introduced in a ultra-high vacuum (UHV) preparation chamber (base pressure in the 10$^{-11}$ mbar range) and annealed at 670-720~K for about 30 minutes.  MoS$_2$ was then deposited on  SiO$_2$ in a dedicated vacuum chamber (base pressure in the $10^{-9}$~mbar range) without exposing the substrate to the atmosphere. MoS$_2$ films were obtained through pulsed laser deposition (PLD) by ablating a bulk MoS$_2$ target (Testbourne Ltd., UK) with a KrF excimer pulsed laser (248~nm wavelength, 20~ns pulse duration). The target was ablated by focused laser pulses emitted at 1 pulse/s rate, and its roto-translational motion was optimized to maximise uniformity of the surface ablation. With a nominal pulse energy of 200~mJ, the fluence on the target averaged to 2~J/cm$^{2}$ per pulse. During the ablation, the SiO$_2$ surface was placed 3~cm away from the target and kept at room temperature (RT). As a result, the number of laser pulses, in the 30-to-120 pulses range, is the sole parameter determining the sample thickness. After deposition, the sample was annealed to 720-810~K for 30 min in UHV to promote the crystallisation of the deposited species.

Raman spectroscopy and photoluminescence measurements were performed ex-situ, employing two distinct InVia Renishaw spectrometers: the first, equipped with 532~nm (2.33~eV) and 660~nm (1.88~eV) laser excitation lines, and the second equipped with 457~nm (2.71~eV) and 514~nm (2.41~eV) laser excitation lines, having a nominal spectral resolution of 3~cm$^{-1}$. We used the 457~nm, 532~nm, and 660~nm excitation lines for Raman spectroscopy measurements, acquiring 3 spectra for each sample, spanning an area of about 1 cm$^2$, and averaging the results for reliable statistics. We used the 514~nm excitation line for photoluminescence measurements. The laser power delivered to the sample was approximately 7~mW during Raman measurements and 1~mW during PL. A 50× objective lens was employed for all measurements. Calibration of the spectrometer was done against the 521~cm$^{-1}$ peak of a Si(100) crystal. Raman modes were fitted using Fityk \cite{wojdyr2010fityk} with linear combinations of Voigt functions after careful background subtraction.

Raman measurements were also performed on exfoliated MoS$_2$ flakes for reference purposes. The monolayer flakes were produced via gold tape exfoliation onto a $200~\mu{m}$ thick SiO$_2$ substrate and analyzed with the same Raman spectrometer and parameters used for the other systems.

Morphology investigation was performed through a field-emission scanning electron microscope (FE-SEM, Zeiss SUPRA 40).  SEM images were acquired by an In-lens detector for secondary electrons,  using an accelerating voltage of 5~kV.

\section*{Results and Discussion}

\subsection{Monolayer production via room temperature PLD on SiO$_2$}

\begin{figure}[htbp]
    \centering    \includegraphics[width=1\textwidth]{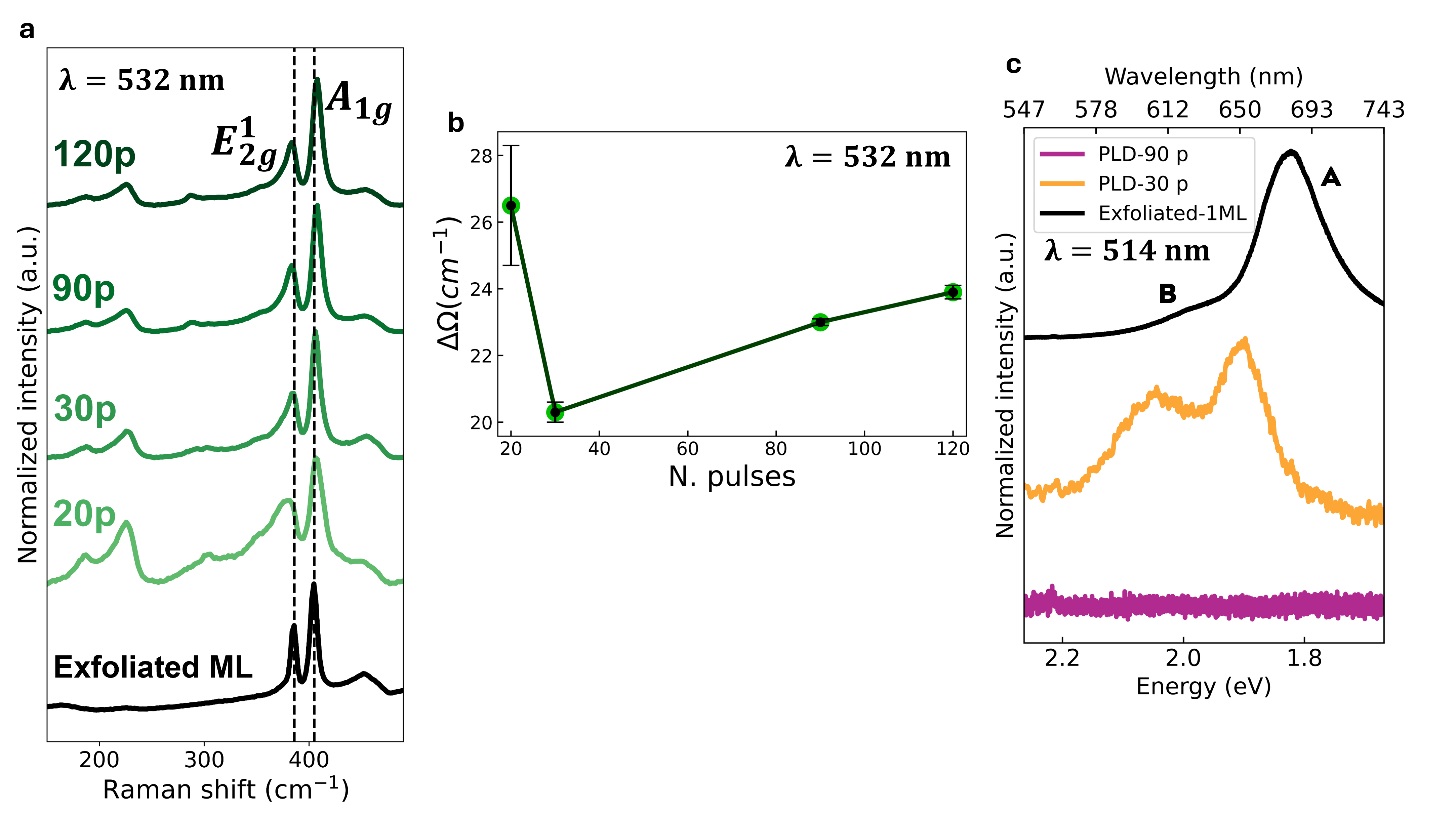}
    \caption{(a) Raman spectra, acquired at 532~nm, of low-dimensional MoS$_2$ samples deposited via PLD on SiO$_2$ at varying number of laser pulses (p) used during the ablation process, from 20 to 120 (green scale curves), compared to the exfoliated monolayer (black curve). The position of E\textsubscript{2g}\textsuperscript{1} and A\textsubscript{1g} first-order Raman peaks for the monolayer exfoliated on SiO$_2$ is reported for reference as black dashed lines. All spectra are normalized to the A\textsubscript{1g} peak. (b) Raman shift difference $\Delta\omega$ between A\textsubscript{1g} and E\textsubscript{2g}\textsuperscript{1} peaks as a function of the number of laser pulses employed during the ablation process. Raman excitation wavelength: 532~nm. (c) Photoluminescence of monolayer (yellow) and multilayer (violet) MoS$_2$ deposited via PLD on SiO$_2$, and of monolayer MoS$_2$ obtained via mechanical esfoliation (black) on SiO$_2$. PL excitation wavelength: 514~nm.    
    }
    \label{fig1}
\end{figure}

2D MoS$_2$ was deposited on SiO$_2$ via room temperature PLD followed by thermal annealing (see Methods for the details), according to a procedure introduced by previous works from our group on 2D MoS$_2$ grown on clean, atomically flat Au(111) and Ag(111) surfaces \cite{tumino2019pulsed,tumino2021hydrophilic}. Figure \ref{fig1}a shows the Raman spectra, acquired with 532~nm excitation, of MoS$_2$ films obtained by PLD using 20, 30, 90 and 120 laser pulses, from now on called "20 p", "30 p", "90 p" and "120 p", compared to the MoS$_2$ monolayer obtained through mechanical exfoliation on SiO$_2$ (see Methods for details). The positions of E\textsubscript{2g}\textsuperscript{1} (386~$cm^{-1}$) and A\textsubscript{1g} (405~$cm^{-1}$) modes for the exfoliated monolayer are reported for reference. The presence of a weakly interacting substrate, i.e., SiO$_2$, allows us to exploit the Raman shift difference $\Delta\omega$ between the two main first-order modes, i.e. E\textsubscript{2g}\textsuperscript{1} and A\textsubscript{1g}, as a reliable metric for determining the number of MoS$_2$ layers \cite{bertrand1991surface,lee2010anomalous,li2012bulk,lee2015raman}.  As shown in Figure \ref{fig1}b, 120 p, 90 p, and 30 p films grown by PLD are associated with a $\Delta\omega$ of 23.9$\pm$0.2 ~$cm^{-1}$, 23$\pm$0.1~$cm^{-1}$ and 20.3$\pm$0.3~$cm^{-1}$, respectively, indicating a progressively decreasing number of MoS$_2$ layers \cite{bertrand1991surface,lee2010anomalous,li2012bulk,lee2015raman}. Notably, an anomalous behaviour is observed for the 20 p sample, whose $\Delta\omega$ increases again. Previous works on high-temperature PLD-grown MoS$_2$ on inert surfaces allow us for assigning the 30 p sample to a monolayer, exhibiting, at 532~nm excitation wavelength, a $\Delta\omega$ of 20.3$\pm$0.3~$cm^{-1}$, with E\textsubscript{2g}\textsuperscript{1} at 385.7~$cm^{-1}$ and A\textsubscript{1g} at 406~$cm^{-1}$ \cite{serna2016large,serrao2015highly,sinha2021study,zhan2011large}, also in agreement with the value obtained for the monolayer exfoliated on SiO$_2$, showing a $\Delta\omega$ of 18.8$\pm$0.2~$cm^{-1}$ (Figure \ref{fig1}a, black curve).

To further confirm that the 30p sample is a monolayer, we performed PL measurements using a 514~nm excitation. 
In Figure \ref{fig1}c, the PL of the 30 p sample (yellow) and 90 p sample (purple) are shown. The A and B excitonic peaks \cite{ellis2011indirect,mak2010atomically,scheuschner2012resonant,mccreary2018and} visible in the 30 p sample, located at 1.89 and 2.1~eV, respectively, are absent in the 90 p sample, confirming the 30 p sample single-layer thickness and the multi-layered structure of the 90 p sample.
Figure \ref{fig1}c also reports the PL of the monolayer obtained via mechanical exfoliation on SiO$_2$ (black curve). The comparison between the PL of the monolayer obtained via PLD and mechanical exfoliation highlights an energy upshift and a variation of the relative intensity of the two contributions of the A and B excitonic peaks from the exfoliated to the PLD-grown sample, possibly due to
the larger defectiveness of the PLD-grown MoS$_2$ film, as shown and discussed in the following.

Beyond the A and B excitons, the presence of higher energy excitonic states, commonly referred to as C excitons, has been recently demonstrated, giving rise to a broad absorption band centred around 2.7~eV  \cite{qiu2013optical,shi2013exciton,li2019slow}. According to many-body calculations of the optical spectrum of MoS$_2$, the C excitons originate from six nearly degenerate exciton states made from transitions between the highest valence band and the first three lowest conduction bands close to $\Gamma$. Unlike the A and B excitons, which are predominantly derived from Mo\textsubscript{d$_{z^2}$} orbitals and exhibit azimuthally symmetric wavefunctions, the C excitons have a mixed orbital character, involving both Mo\textsubscript{d$_{z^2}$} orbitals and S p\textsubscript{x} and p\textsubscript{y} states \cite{sekine1980resonance,qiu2013optical,carvalho2016erratum}.

As demonstrated by McCreary et al. \cite{mccreary2018and}, excitons lifetimes can be largely influenced by the presence of defects, which generate localized electronic states within the gap acting as trapping sites for excitons, resulting in progressively shorter non-radiative recombination times \cite{mccreary2018and,burns2020controlling}. In high-quality exfoliated MoS$_2$ monolayer samples, i.e., with low defect densities, the PL spectrum is dominated by the A exciton, while the B exciton appears as a much weaker feature, yielding a small B/A intensity ratio. When the system disorder increases, non-radiative recombination channels associated with defects significantly shorten the lifetime of the A exciton. In contrast, the B exciton exhibits an intrinsically much shorter lifetime, primarily governed by rapid intraband relaxation to the A exciton, making its recombination dynamics comparatively insensitive to additional defect-induced scattering , resulting into a larger B/A ratio \cite{mccreary2018and,burns2020controlling,qiu2013optical}. Differently from A and B bound excitons, localized at the band edge, C excitons derive from MoS$_2$ band nesting region, resulting highly delocalized in the k-space \cite{qiu2013optical,shi2013exciton,li2019slow,qu2024defect}. This translates in different relaxation dynamics for the C excitons, with ultrafast intraband relaxation (< 500 fs) to A-B excitons region \cite{shi2013exciton}, possibly slowed down by intervalley transfer mechanisms \cite{li2019slow}.
As a result, an increasing defect density is expected to suppress the A-exciton PL intensity more strongly than that of the B and C excitons. This is exemplified by Figure \ref{fig1}c, where the B/A excitonic peak ratio assumes a value of 0.17 in the monolayer obtained via mechanical exfoliation and of 0.7 in the monolayer obtained via PLD, indicating the much larger defectiveness of the PLD-grown monolayer.

The variation of the $\Delta\omega$ value with the number of laser pulses (Figure \ref{fig1}b) allowed us also to estimate a 3-layer configuration to the 90 p sample and a 4-layer configuration to the 120 p sample \cite{bertrand1991surface,lee2010anomalous,li2012bulk,lee2015raman,serna2016large,serrao2015highly,sinha2021study,zhan2011large}. On the other hand, an unexpected behaviour is observed in correspondence of the 20 p sample, exhibiting a sharp increase in the value of $\Delta\omega$ up to 26.5$\pm$1.8~$cm^{-1}$. Scanning electron microscopy (SEM) measurements show a huge morphological variation between the 20 p and 30 p sample, with the former characterized by islands separated by exposed regions of the SiO$_2$ surface and the latter by a highly uniform coverage (see Figure S1). This behaviour can be related to the different degree of interaction between MoS$_2$ and the substrate surface with respect to the growth on metals such as Au(111) or Ag(111). Indeed, MoS$_2$ is known for interacting strongly with metals \cite{tumino2025surface, tumino2021hydrophilic}, while its behaviour resembles the freestanding one when exfoliated or grown on SiO$_2$ \cite{bertrand1991surface,lee2010anomalous,li2012bulk,lee2015raman}. Combining the information gained by SEM images with the evolution of the $\Delta\omega$ value as a function of the MoS$_2$ samples thickness, we hypothesize that, on SiO$_2$, when the amount of deposited material is below an equivalent monolayer, i.e. less than 30 laser pulses, MoS$_2$ tends to preferentially undergo self-interaction with the MoS$_2$ species that already reached the surface, piling up into multilayered islands, thus leaving exposed SiO$_2$ regions. Conversely, on metals, the favoured interaction between MoS$_2$ and the surface allows for the direct formation of contiguous monolayer islands that, upon annealing, merge into a uniform single-layer-thick film. On the other hand, increasing the amount of material deposited on SiO$_2$ up to the monolayer threshold, i.e. 30 laser pulses in our setup, upon annealing-enhanced surface mobility, leads to the formation of a continuous single layer.
Based on previous work on PLD-grown MoS$_2$ on Au(111) \cite{tumino2019pulsed}, we hypothesize that, also on SiO$_2$, the monolayer is constituted by crystal domains having an extension of tens of nanometers. On the other hand, when the amount of material deposited on SiO$_2$ overcomes the monolayer threshold, i.e. 30 laser pulses in our setup, multilayer MoS$_2$ is obtained upon annealing, similar to what is observed on metal surfaces.

By means of Raman spectroscopy, we also investigated the stability in atmospheric conditions of 2D MoS$_2$ synthesized by PLD (see Figure S2). Both the monolayer and the 4-layer sample preserve their spectral features up to 20 months after the deposition (Figure S2 a-b), with a non-negligible decrease in the intensity of the E\textsubscript{2g}\textsuperscript{1}
and A\textsubscript{1g} first-order Raman modes, symptomatic
of the samples degradation (Figure S2 c).

\subsection{Defects in 2D MoS$_2$}

\begin{figure}[htbp]
    \centering    \includegraphics[width=1\textwidth]{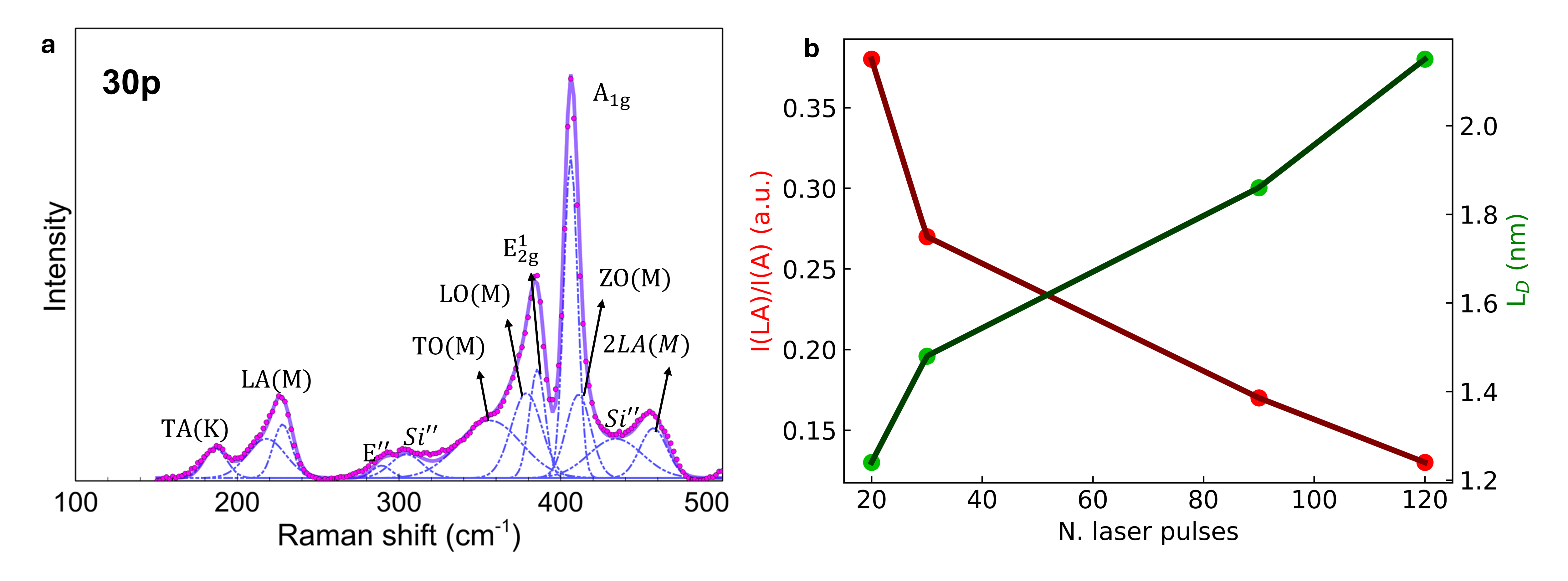}
    \caption{(a) Example of Raman spectrum of MoS$_2$ monolayer grown via PLD on SiO$_2$ fitted by Voigt functions. Magenta circles represent the raw data, the dashed blue lines are the deconvoluted components of the Raman peaks, and the violet solid line represents the resulting fit. (b) Intensity ratio between LA(M) and A\textsubscript{1g} peaks (red) and interdefect distance L$_D$ (green) as a function of the number of laser pulses employed during the ablation process. Raman excitation wavelength: 532~nm.}
    \label{defect_Mignuzzi}
\end{figure}

Figure \ref{defect_Mignuzzi}a shows the fitting of the Raman modes, acquired at 532 nm, of MoS$_2$ monolayer grown by PLD on SiO$_2$. Besides the E\textsubscript{2g}\textsuperscript{1} and A\textsubscript{1g} first-order Raman peaks, in the spectral range from 188 to 454 cm$^{-1}$ it is possible to recognize several other spectral features, associated with second-order (Si'', 2LA(M)) and disorder-activated (TA(K), LA(M), TO(M), LO(M), ZO(M)) vibrational modes involving phonons at the edge of the Brillouin zone \cite{mignuzzi2015effect,uchinokura1974critical}. In particular, it is important to notice that the defect-activated peaks visible in all films obtained through our PLD-based approach are absent in the monolayer obtained through mechanical exfoliation (see Figure \ref{fig1}a). The assignment of such peaks for MoS$_2$ monolayer is established in the literature \cite{mignuzzi2015effect,uchinokura1974critical,wu2016raman}, and it is summarized in Table 1. Among defect-activated modes, it is worth mentioning the longitudinal acoustic phonon at the M point, commonly denoted as LA(M), at about 227~$cm^{-1}$. This mode arises from the near‐flat dispersion of the longitudinal acoustic branch at the M point of the Brillouin-zone, which determines a local enhancement of the vibrational density of states at the zone-edge \cite{mignuzzi2015effect,frey1999raman,molina2011phonons}.

Mignuzzi et al. \cite{mignuzzi2015effect} investigated the effect of ion bombardment-induced defects on the Raman spectrum of a MoS$_2$ monolayer mechanically exfoliated on SiO$_2$, and found a particularly strong correlation between structural defects and the intensity of the LA(M) mode, which increases for decreasing interdefect distance L\textsubscript{d} (i.e., increasing defect density). In particular, they found that the intensity of the LA(M) peak, normalized to the E\textsubscript{2g}\textsuperscript{1} or A\textsubscript{1g} peak intensity, is inversely proportional to L\textsubscript{d}\textsuperscript{2}, according to the following relation: $\frac{I(\text{LA})}{I(X)} = \frac{C(X)}{L_d^2}$, being X = E\textsubscript{2g}\textsuperscript{1} or A\textsubscript{1g}, and C a defect-related scattering cross-section coefficient, whose value depends on the employed Raman probe. Using 532 nm as excitation wavelength, they extracted the values C(E\textsubscript{2g}\textsuperscript{1}) = 1.11 ± 0.08~nm\textsuperscript{2} and C(A\textsubscript{1g}) = 0.59 ± 0.03~nm\textsuperscript{2}.

\begin{table}[htbp]
\centering
\begin{tabular}{|c|c|} 
\hline
\multicolumn{2}{|c|}{\textbf{Second-order and Defect-activated peaks in MoS$_2$ monolayer}} \\
\hline
Frequency ($cm^{-1}$) & Assignment \\
\hline
188 & TA(K) \\
215 & - \\
227 & LA(M) \\
303 & Si'' \\
357 & TO(M) \\
377 & LO(M) \\
412 & ZO(M) \\
434 & Si'' \\
454 & 2LA(M) \\
\hline
\end{tabular}
\caption{Tabulated Raman shift and corresponding mode assignment reported in the literature for second-order and defect-activated peaks in MoS$_2$ monolayer. Second-order Si'' Raman mode of silicon is taken from Ref. \citenum{richter1981one}. Defect-activated peaks induced on MoS$_2$ monolayer exfoliated on SiO$_2$ by ion-bombardment are taken from Ref. \citenum{mignuzzi2015effect}. TA stays for "transverse acoustic", LA for "longitudinal acoustic", TO for "transverse optical", LO for "longitudinal optical", and ZO for "out-of-plane optical".}
\end{table}
\vspace{20pt}

Interestingly, a similar behaviour is found for the intensity ratio of the D and the G peaks in graphene, widely employed to determine its defectiveness \cite{pollard2014quantitative}. Mignuzzi's model has also been employed for assessing the average interdefect distance in MoS$_2$ monolayer samples grown via CVD and exposed to hydrogen and oxygen plasmas \cite{stanford2019lithographically}. In the same way, we employed it to estimate the average interdefect distance in our MoS$_2$ samples grown by PLD on SiO$_2$, according to the relation: $\frac{I(\text{LA(M)})}{I(A_{1g})} = \frac{0.59 \pm 0.03 nm^2}{L_d^2}$. In Figure \ref{defect_Mignuzzi}b, it can be observed that for decreasing number of deposited layers, the intensity ratio between LA(M) and A\textsubscript{1g} increases, while the value of L$_D$ decreases from 2.15~nm $\pm$ 0.05 in the quadrilayer to 1.86~nm $\pm$ 0.05 in trilayer, 1.48~nm $\pm$ 0.04 in the monolayer and, finally, 1.24~nm $\pm$ 0.03 in the 20 p sample, indicating a progressively higher defect density. This thickness-dependent defectiveness is also confirmed by the increase, from the 120 p to the 20 p samples, of the FWHM of both E\textsubscript{2g}\textsuperscript{1} peak, from 8-9~$cm^{-1}$ to 21~$cm^{-1}$, and the A\textsubscript{1g} peak, from 9~$cm^{-1}$ to 13~$cm^{-1}$ \cite{mignuzzi2015effect}, extracted from the data reported in Figure \ref{fig1}a. 
The obtained L$_D$ value, in the order of a few nanometers, supports our hypothesis on the average size of the crystal domains constituting MoS$_2$ monolayer grown by PLD on SiO$_2$, mentioned in the previous section. Notably, it is also in reasonable agreement with the one reported by Mignuzzi et al. \cite{mignuzzi2015effect} for the MoS$_2$ monolayer exfoliated on SiO$_2$ after its exposure to a density of ions of about 0.39~nm$^{-2}$. As expected, in analogy with MoS$_2$ films grown on Au(111) and Ag(111) \cite{tumino2019pulsed,tumino2021hydrophilic}, MoS$_2$ films obtained through PLD on SiO$_2$ are nanocrystalline, with a large density of grain boundaries, resulting much more defective than those obtained via mechanical exfoliation, which, in the pristine state, do not exhibit any defect-activated peak (Figure \ref{fig1}a, black curve).

\subsection{Multiwavelength Raman investigation of 2D MoS$_2$}

\begin{figure}[htbp]
    \centering    \includegraphics[width=1\textwidth]{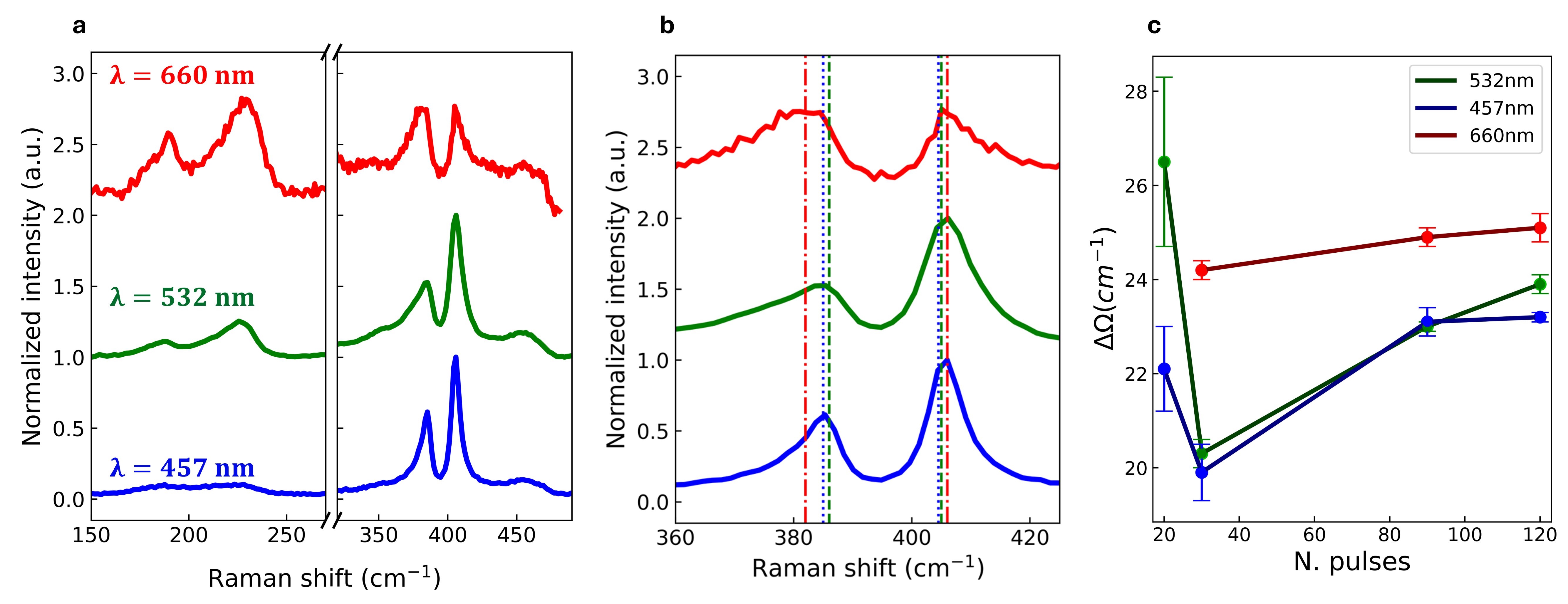}
    \caption{(a) Raman spectra, acquired at 532~nm (green), 457~nm (blue), and 660~nm (red), of MoS$_2$ monolayer deposited via PLD on SiO$_2$. (b) Close-up of (a) in the Raman shift range from 360~cm$^{-1}$ to 425~cm$^{-1}$.  Dashed vertical lines: positions of the E\textsubscript{2g}\textsuperscript{1} and A\textsubscript{1g} peaks in monolayer MoS$_2$ exfoliated on SiO$_2$, acquired at 532~nm (green), 457~nm (blue) and 660~nm (red). (c) Raman shift difference $\Delta\omega$ between A\textsubscript{1g} and E\textsubscript{2g}\textsuperscript{1} peaks as a function of the number of laser pulses employed during the ablation process; Raman excitation wavelengths: 532~nm (green), 457~nm (blue) and 660~nm (red).}
    \label{monolayer_different_lasers}
\end{figure}

We investigated the excitation wavelength-dependence of the Raman response of PLD-grown MoS$_2$ monolayer (Figure \ref{monolayer_different_lasers}a-b) and multilayer (Figure S3) on SiO$_2$. Raman spectra acquired at 660~nm (1.88~eV) excitation reveal a substantially stronger contribution of first- and second-order Raman features than those at 457~nm (2.71~eV) and 532~nm (2.33~eV), suggesting the existence of pronounced resonance-related effects. 

Beyond intensity effects, resonance conditions also affect first-order phonon energies. While the E\textsubscript{2g}\textsuperscript{1} and A\textsubscript{1g} peaks positions remain unchanged within instrumental uncertainty under 457 nm and 532 nm excitation, resulting in a $\Delta\omega$ of 20.3$\pm$0.3~cm$^{-1}$, excitation at 660 nm induces a downshift (about 4~cm$^{-1}$) of the E\textsubscript{2g}\textsuperscript{1} mode, increasing the value of $\Delta\omega$ to 24.2$\pm$0.2~cm$^{-1}$ (Figure \ref{monolayer_different_lasers}c).
An identical excitation-dependent trend, with overlapping peak positions, is observed in mechanically exfoliated monolayers on SiO$_2$, whose Raman fingerprints at 457~nm, 532~nm and 660~nm are reported as vertical dashed lines in Figure \ref{monolayer_different_lasers}b, indicating that PLD-grown and exfoliated samples share equivalent intrinsic vibrational properties.
Therefore, the observed spectral differences arise from resonant excitation conditions rather than structural differences between the two systems.

Notably, the same trend observed at 532~nm for the $\Delta\omega$ as a function of the number of deposited layers can also be observed at 457~nm and 660~nm, as shown in Figure \ref{monolayer_different_lasers}c. The value of $\Delta\omega$ corresponding to the 20 p sample using the 660 nm excitation wavelength is not present due to
the too small signal-to-noise ratio of the Raman spectra, which did not allow for a proper fit.

Having established the coherence of the first-order Raman response between the PLD-grown and exfoliated MoS$_2$ monolayers at different excitation wavelengths, we further analyze the excitation dependence of second-order scattering processes, demonstrating how strongly the vibrational properties of MoS$_2$ monolayers are affected by resonant excitation conditions. As already mentioned above, PL spectra of MoS$_2$ monolayers are characterized by the presence of A and B excitonic features, located at 1.89 and 2.1 eV, respectively.

\begin{figure}[htbp]
    \centering    \includegraphics[width=1\textwidth]{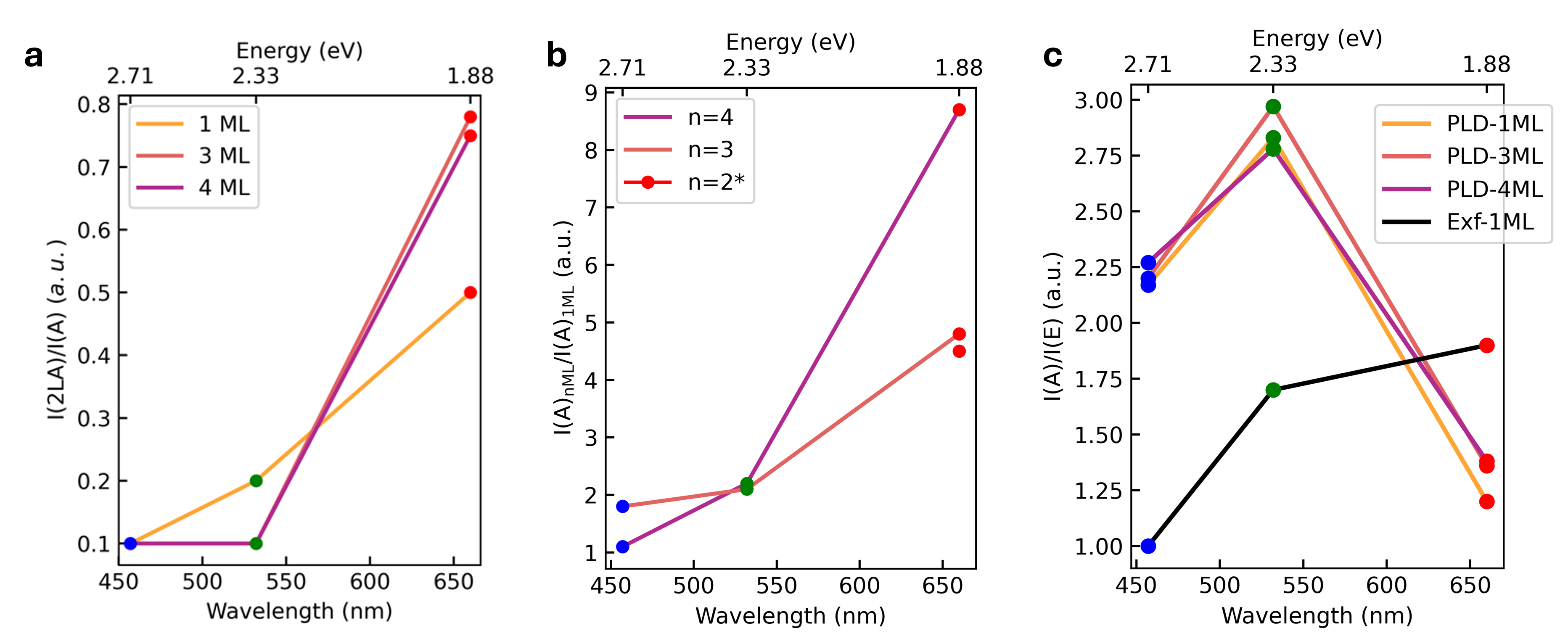}
    \caption{(a) Intensity ratio between 2LA(M) and A\textsubscript{1g} peaks in low-dimensional MoS$_2$ samples deposited via PLD on SiO$_2$ as a function of the excitation wavelength: 532~nm (green), 457~nm (blue), and 660~nm (red). (b) A\textsubscript{1g} peak intensity ratio in trilayer and quadrilayer MoS$_2$ with respect to monolayer MoS$_2$ grown by PLD on SiO$_2$ as function of the excitation wavelength. The value of the intensity ratio for the bilayer MoS$_2$ is taken from Ref. \citenum{scheuschner2012resonant} as reference. (c) Intensity ratio between A\textsubscript{1g} and E\textsubscript{2g}\textsuperscript{1} peaks in monolayer MoS$_2$ obtained via PLD (yellow) and mechanical exfoliation (black) on SiO$_2$ as a function of the excitation wavelength.}
    \label{exciton}
\end{figure}

Figure \ref{exciton}a reports the 2LA(M)/A\textsubscript{1g} intensity ratio for mono-, tri-, and four-layer MoS$_2$ grown by PLD on SiO$_2$ as a function of the excitation wavelength. It can be observed that, independently of the film thickness, the 2LA(M)/A\textsubscript{1g} intensity ratio undergoes an increase from 457~nm to 532~nm and 660~nm excitation, becoming particularly large close to 660~nm.
This behaviour is consistent with the well-known resonance enhancement of overtone modes in MoS$_2$, attributed to a double-resonance Raman process involving two LA(M) phonons with opposite momenta \cite{placidi2015multiwavelength,zhang2015phonon,mignuzzi2015effect}, analogous to the defect-activated D and 2D bands in graphene.

This explains the increase in the 2LA(M)/A\textsubscript{1g} intensity ratio in correspondence of 660~nm (1.88~eV), i.e. in resonance with the A exciton \cite{zhang2015phonon,placidi2015multiwavelength}. Notably, the 2LA(M)/A\textsubscript{1g} intensity ratio in correspondence of the 660~nm excitation wavelength increases from the mono- to the four-layer sample, assuming substantially the same values for the 3- and 4-layer configurations, suggesting a thickness-dependent efficiency of resonant scattering involving the A exciton and the 2LA(M) phonons in 2D MoS$_2$.

We also observe the existence of a thickness-dependent exciton-phonon coupling between A exciton and A\textsubscript{1g} Raman mode in 2D MoS$_2$, as it is shown in Figure \ref{exciton}b, reporting the A\textsubscript{1g}(n-layer)/A\textsubscript{1g}(monolayer) intensity ratio (for n = 3 and 4) as a function of the excitation wavelength. It can be noticed a sharp increase in the value of the A\textsubscript{1g}(n-layer)/A\textsubscript{1g}(monolayer) intensity ratio under A-exciton resonance, i.e., 660~nm (1.88~eV), which becomes larger as the number of MoS$_2$ layers increases. Notably, the 2LA(M) mode has a stronger thickness-dependent coupling with A exciton with respect to the A\textsubscript{1g} mode. Indeed, under A-exciton resonance conditions, the $2LA(M)_{4-layer}$ peak intensity is 13 times larger than that of $2LA(M)_{monolayer}$, while the A\textsubscript{1g}(4-layer) peak intensity is 9 times larger than that of A\textsubscript{1g}(monolayer).
Our hypothesized thickness-dependent exciton-phonon coupling in 2D MoS$_2$is also supported by previous works by Scheuschner et al. \cite{scheuschner2012resonant, scheuschner2015interlayer} on exfoliated MoS$_2$ samples.

Further insight into the interplay between excitonic transitions and lattice vibrations in MoS$_2$ monolayer emerges from the excitation wavelength dependence of the A\textsubscript{1g}/E\textsubscript{2g}\textsuperscript{1} intensity ratio (Figure \ref{exciton}c). In particular, we experimentally show the theoretically predicted \cite{qiu2013optical} effects of the symmetry-dependent coupling between the out-of-plane A\textsubscript{1g} with the orbitals associated with A and B excitons, and between the in-plane E\textsubscript{2g}\textsuperscript{1} phonons and the orbitals associated with the C excitons in 2D MoS$_2$, which was previously discussed also by Carvalho et al. \cite{carvalho2016erratum}; this coupling affects the relative A/E ratio at different excitation wavelengths.

Considering the exfoliated MoS$_2$ monolayer (Figure \ref{exciton}c, black curve), when compared to the off-resonance excitation (i.e. 532~nm), the A\textsubscript{1g}/E\textsubscript{2g}\textsuperscript{1} ratio decreases under 457~nm excitation (C-exciton resonance), due to the preferential enhancement of the E\textsubscript{2g}\textsuperscript{1} mode, and increases under 660~nm excitation (A-exciton resonance), reflecting selective enhancement of the out-of-plane A\textsubscript{1g} phonon. This behaviour is consistent with the distinct orbital symmetries predicted for A, B and C excitons in 2D MoS$_2$ described in the previous section.

While the PLD-grown MoS$_2$ monolayer (Figure \ref{exciton}c, yellow curve) reproduces the resonance behaviour at 457 nm (C exciton), a markedly different response is observed under 660~nm excitation (A-exciton resonance): instead of the expected enhancement with respect to the off-resonance excitation (i.e. 532 nm), the A\textsubscript{1g}/E\textsubscript{2g}\textsuperscript{1} ratio exhibits a pronounced suppression. This trend is also coherent for the 3-layer and 4-layer MoS$_2$ samples (Figure \ref{exciton}c, orange and violet curves). Notably, the suppression of the intensity ratio under 660 nm excitation is smaller for the 3- and 4-layer samples than for the monolayer, confirming the thickness-dependent exciton-phonon coupling observed in Figure \ref{exciton}a-b. 
We attribute this behaviour to the higher defect density of the PLD-grown film with respect to the one obtained via mechanical exfoliation. As mentioned in the previous section (Figure \ref{fig1}c), in defective samples, the A-exciton PL intensity is quenched more strongly than that of the B and C excitons. As a consequence, in light of our symmetry-dependent exciton-phonon coupling interpretation, when in resonance with A exciton, the A\textsubscript{1g} mode scattering cross section is largely reduced, resulting in the observed large decrease in the A\textsubscript{1g}/E\textsubscript{2g}\textsuperscript{1} ratio (Figure \ref{exciton}c). 

\section*{Conclusions}
This work demonstrates the possibility to employ a room temperature PLD-based approach for the obtainment of 2D MoS$_2$ samples with finely tunable thickness, down to the monolayer, on electronics-friendly SiO$_2$ substrates. Through the combination of Raman spectroscopy and SEM imaging, we delve into their structural and vibrational properties. The trend of the Raman shift difference ($\Delta\omega$) as a function of the number of deposited layers, combined with the estimation of the interdefect distance by means of the disorder-activated Raman modes, allowed us to study the growth of 2D MoS$_2$ on SiO$_2$ as a function of the number of laser pulses employed during the PLD process. Through the combination of resonant Raman and photoluminescence spectroscopy, we observe a thickness- and symmetry-dependent exciton-phonon coupling in 2D MoS$_2$, resulting in a stronger coupling of the A excitons with the out-of-plane A\textsubscript{1g} mode and a stronger coupling of the C excitons with the E\textsubscript{2g}\textsuperscript{1} mode. The comparison between samples obtained via PLD and mechanical exfoliation revealed that defects can affect exciton-phonon coupling in PLD-grown 2D MoS$_2$.

\section*{Data availability}
The data supporting the findings of this study are available in the Article and its Supplementary Information. Additional data are available from the corresponding authors on request.

\section*{Author contributions}
A.C. synthesized MoS$_2$ samples via PLD, performed Raman spectroscopy, PL measurements and SEM imaging. F.T. conducted Raman and PL measurements on mechanically exfoliated MoS$_2$ samples. A.C. analyzed the data and wrote the initial draft. V.R., C.S.C., and A.L.B. contributed to the supervision and discussion of results. All authors contributed to the revision and final discussion of the manuscript.

\section*{Conflicts of interest}
There are no conflicts of interest to declare.

\section*{Acknowledgments}
A.C., V.R., A.L.B. and C.S.C acknowledge funding by: - Funder: project funded under the National Recovery and Resilience Plan (NRRP), Mission 4 Component 2 Investment 1.3 - Call for tender No. 341 of 15.03.2022 of Ministero dell’Universita' e della Ricerca (MUR); funded by the European Union NextGenerationEU - Award Number: project code PE0000021, Concession Decree No. 1561 of 11.10.2022 adopted by Ministero dell’Università e della Ricerca (MUR), CUP D43C22003090001, Project title “Network 4 Energy Sustainable Transition NEST”.

\bibliography{biblio}

\normalsize

\end{document}


\maketitle
\clearpage
\section{Scanning electron microscopy}

\begin{figure}[htbp]
\centering
\includegraphics[width=0.9\textwidth]{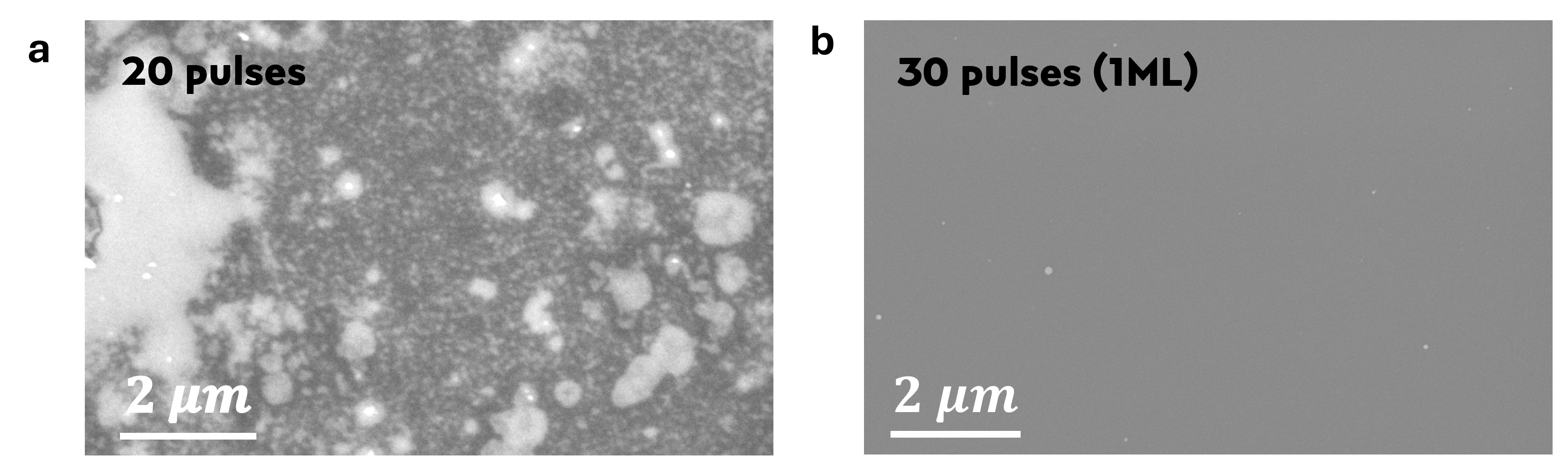}
\caption{SEM top-view image of MoS$_2$ bulk islands (a) and monolayer (b) on SiO$_2$/Si.
}\label{SEM}
\end{figure}

We performed scanning electron microscopy on the 20p and 30p samples, whose images are reported in Figure \ref{SEM}. Figure \ref{SEM}a shows separated islands of MoS$_2$ (dark gray) that leave the underneath SiO$_2$ substrate exposed and well visible by optical contrast (light gray). Instead, Figure \ref{SEM}b shows a highly uniform surface, corresponding to the MoS$_2$ monolayer, without any exposed portion of the underneath SiO$_2$.

\newpage
\section{Stability in atmospheric conditions}

\begin{figure}[htbp]
\centering
\includegraphics[width=1\textwidth]{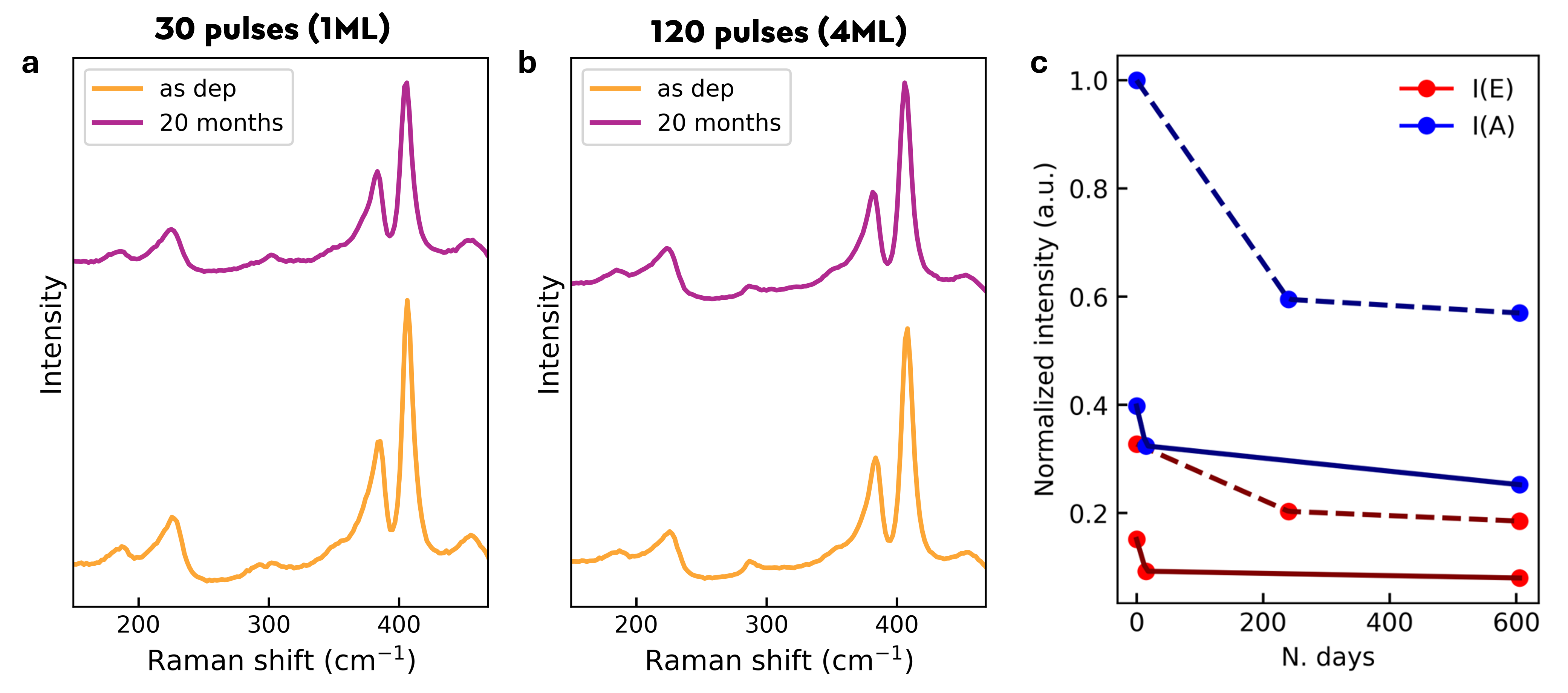}
\caption{Raman spectra of MoS$_2$ monolayer (a) and 4-layer (b) samples acquired just after the deposition (as dep) and after 20 months of exposure to the atmospheric conditions. Raman excitation wavelength: 532~nm. (c) Intensity of E\textsubscript{2g}\textsuperscript{1} and A\textsubscript{1g} peaks as a function of the number of days under atmospheric exposure for the monolayer (solid lines) and 4-layer (dashed lines) sample. Raman excitation wavelength: 532~nm. 
}\label{stability_atm}
\end{figure}

We performed ex situ Raman spectroscopy, with 532 nm excitation wavelength, on the MoS$_2$ monolayer (30 p) and 4-layer (120 p) samples after 20 months of exposure to the atmospheric conditions, and compared the results with the spectra of the corresponding samples taken just after their synthesis, as shown in Figure \ref{stability_atm}. With respect to the as-deposited system, the indication of the sample degradation after 20 months of exposure to the atmospheric conditions is consistent between the mono- and 4-layer systems. Indeed, the monolayer shows a decrease in the intensity of the E\textsubscript{2g}\textsuperscript{1}
and A\textsubscript{1g} modes of 1.9 and 1.6 times, respectively, while the 4-layer of 1.8 for both peaks. Notably, the decrease in the intensity of the first-order Raman modes observed for the monolayer is much steeper in the first 15 days of exposure than in the remaining time, till 20 months. This suggests that, probably, the degradation of the sample occurs faster in the first weeks of exposure to the atmospheric conditions and then it progressively slows down.

\newpage
\section{3-layer and 4-layer multiwavelength Raman characterization}
\begin{figure}[htbp]
\centering
\includegraphics[width=1\textwidth]{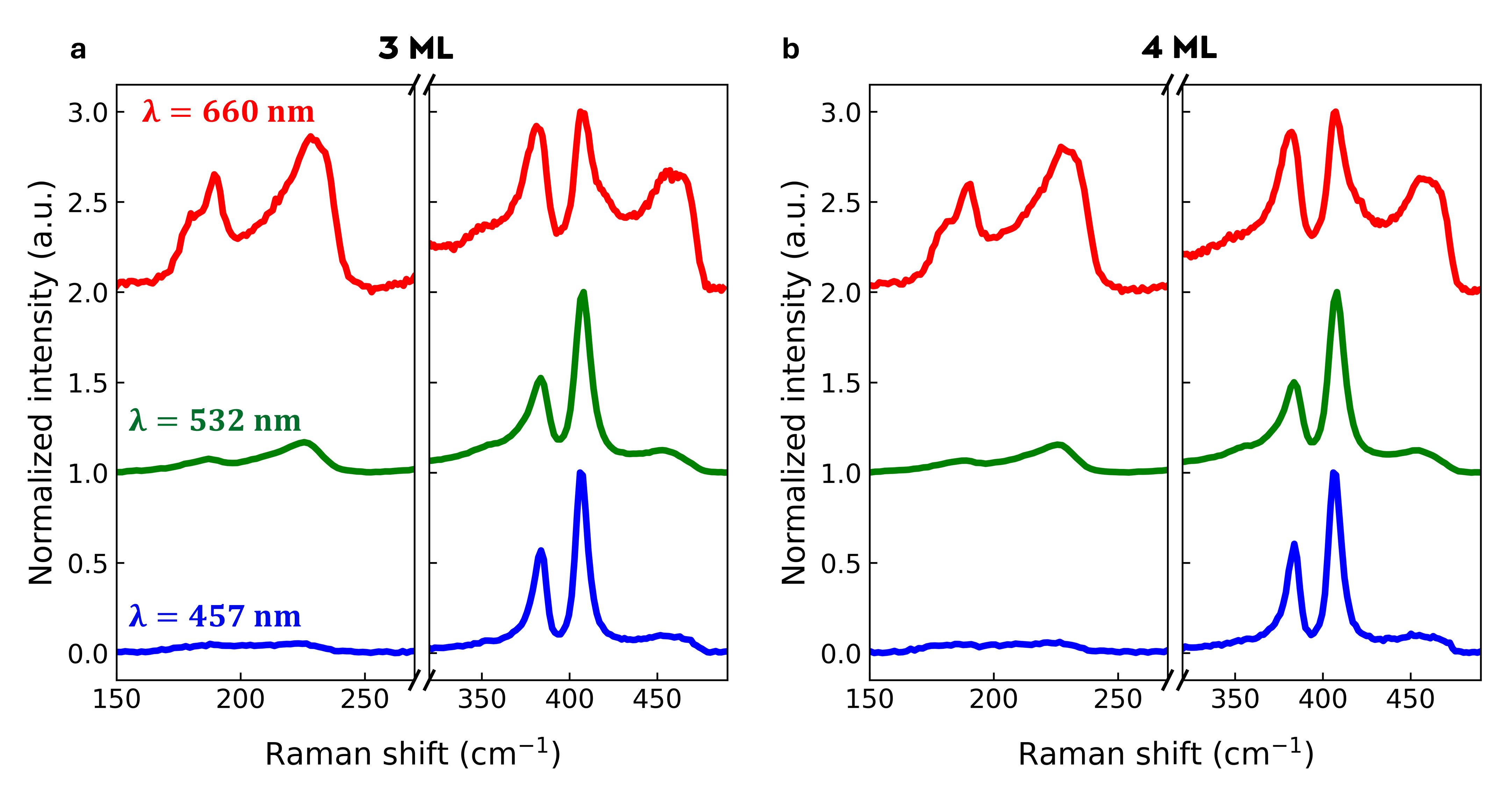}
\caption{(a) Raman spectra, acquired at 532~nm (green), 457~nm (blue), and 660~nm (red), of 3-layer MoS$_2$ deposited via PLD on SiO$_2$. (b) Raman spectra, acquired at 532~nm (green), 457~nm (blue), and 660~nm (red), of 4-layer MoS$_2$ deposited via PLD on SiO$_2$.
}\label{3L-4L}
\end{figure}

Figure \ref{3L-4L} shows the Raman spectra acquired at 532~nm (green), 457~nm (blue), and 660~nm (red), of 3- and 4-layer MoS$_2$ deposited via PLD on SiO$_2$.

